# Nondegenerate parametric down conversion in coherently prepared two-level atomic gas

[1]Gevorg Muradyan[a], Atom Zh. Muradyan[b]

[a]Dep. of Physics, TU Kaiserslautern, Erwin-Schrödinger str., 67663 Kaiserslautern, Germany
[b]Dep. of Physics, Yerevan State University, 1 A. Manookian 375025, Yerevan, Armenia

We describe parametric down conversion process in a two-level atomic gas, where the atoms are in a superposition state of relevant energy levels. This superposition results in splitting of the phase matching condition into three different conditions. Another, more important, peculiarity of the system under discussion is the nonsaturability of amplification coefficients with increasing pump wave intensity, under "sideband" generation conditions.

## 1. Introduction

Parametric down conversion (PDC) [1] is a three-wave interaction process where a pump photon of frequency $\omega_p$ "splits" into two, signal and idler, photons with lower frequencies $\omega_s$ and $\omega_i$ respectively. To break parity-conservation selection rule, so that it is not forbidden for one photon to decay into two photons, a medium with a lack of inversion symmetry is needed. In experimental setups, this request often is satisfied by employing some non-centrosymmetric crystals, from which $BBO$, $LiIO_3$, and $KNbO_3$ are the best candidates [2]. Note, that the non-classical nature of radiation, such as the sub-Poissonian statistics, the "squeezing" process and the quantum entanglement [3], is exhibited most expressively far below the threshold of generation, in a range of the so called spontaneous PDC.

In recent years the studies of generation of paired photons in quantum dots [4] and atomic gases [5] are receiving serious attention too. The discrete spectra of these ensembles, realized in optical cavities, can be exploited to provide wide absorption-free regions, resonant enhancement in the nonlinear susceptibilities and a narrow emission spectrum. The efficiency of conversion in atomic gases is strongly restricted due to the well-defined parity properties of atomic states. Of course, this is an essential drawback, if the purpose is to construct a radiation source of a new frequency. However, if the quest is to reveal quantum characteristics of radiation, it is much more preferable to detect the separated biphotons directly from the source of generation, instead of using an attenuator in front of a bright radiation source. Also, it is highly desirable to have feasible control of smooth and large scale variation of parameters relevant to the generation process. All these features are inherent to the gaseous medium.

This paper aims to continue and extend the study of PDC in atomic two-level medium for a more general situation, when the conversion is originated from arbitrary superposition of the dressed (in the pump field) atomic states. Bare atomic states are taken of different parity, allowing single-photon transition between the levels in dipole approximation. It will be shown that the phase matching condition for PDC process in this scheme is split into the three conditions (it reminds the Mollow's splitting in case of one-photon luminescence). The main result of the paper is the nonsaturable nature of amplification coefficient in dependence of pump field intensity under both, referred to as "sideband", conditions of parametric generation.

---
[1] Corresponding author E-mail: gmurad@physik.uni-kl.de

## 2. PDC in a medium of two-level atoms prepared in a superposition of dressed states

To address the problem we will use the semiclassical theory of interaction of radiation and two-level medium (the energy separation between the upper and lower internal atomic states $|2\rangle$ and $|1\rangle$ is $\hbar\omega_0$). This will suffice for calculating the presenting interest coefficient of amplification of the PDC process. The spin of the relevant to optical transition electron and the possible sublevel structure of energy levels are not taken into account. The atomic gas is taken as an ideal one and each atom interacts with the radiation field individually. Hence, the field-matter interaction entails a simple substitution $\hat{\mathbf{p}} \to \hat{\mathbf{p}} - (e/c)\mathbf{A}(\mathbf{r},t)$ in $\hat{H}_0$, the free atom Hamiltonian. We write the vector potential as $\mathbf{A}(\mathbf{r},t) = \mathbf{A}_p(z,t) + \mathbf{A}_w(\mathbf{r},t)$, where the pump-part $\mathbf{A}_p(\mathbf{r},t)$ propagates through the medium without intensity changes. The weak field $\mathbf{A}_w(\mathbf{r},t)$ ($\mathbf{A}_w(\mathbf{r},t) \ll \mathbf{A}_p(z,t)$) represents a superposition of signal and idler fields, which amplify while propagating through the medium.

We start with the wave equation

$$\left(\nabla^2 - \frac{1}{c^2}\frac{\partial^2}{\partial t^2}\right)\mathbf{A}_w(\mathbf{r},t) = -\frac{4\pi n}{c}\frac{\partial}{\partial t}\langle\hat{\mathbf{d}}\rangle_w, \qquad (1)$$

where n is the atomic density and $\langle\hat{\mathbf{d}}\rangle_w$ denotes the atom dipole moment induced by the signal and idler fields. We assume that the pump, as well as the signal and idler fields are monochromatic with slowly varying envelopes:

$$\mathbf{A}_w(\mathbf{r},t) = \mathbf{A}_s \exp(i\mathbf{k}_s\mathbf{r} - i\omega_s t) + \mathbf{A}_i \exp(i\mathbf{k}_i\mathbf{r} - i\omega_i t) + \text{c.c.}, \qquad (2)$$

so that the second order derivatives can be omitted. The frequency of pump field is close to the transition frequency connecting two atomic levels, and the detuning from the resonance is greater than the optical line broadening.

To determine $\langle\hat{\mathbf{d}}\rangle_w$ in the right hand side of Eq.(1), one has to find the state vector from Schrödinger equation and carry out quantum mechanical averaging of the operator $\hat{\mathbf{d}}$. What we are concerned with is the electric dipole transition for the pump field and thus for the Hamiltonian of the interacting atom can write

$$\hat{H} = \hat{H}_0 - \hat{\mathbf{d}}\mathbf{E}_p(z,t) + \hat{V}_{weak}(\mathbf{r}_e, t). \qquad (3)$$

Here $\hat{\mathbf{d}}$ is the dipole moment operator and $\mathbf{E}_p(z,t)$ is electric vector of the pump field evaluated at point $z$, position of the atomic center of mass (a.c.m) along the pump field propagation axis. The weak field interaction potential is

$$\hat{V}_{weak}(\mathbf{r}_e, t) = -\frac{e}{c}\mathbf{A}_w(\mathbf{r}_e,t)\hat{\mathbf{p}} - \frac{e^2}{mc^2}\mathbf{A}_p(z_e,t)\mathbf{A}_w(\mathbf{r}_e,t), \qquad (4)$$

and later on is regarded as a perturbation. Note, that both, pump and weak fields, are evaluated at the optical electron's coordinate $\mathbf{r}_e$, and the term proportional to $\mathbf{A}_w^2(\mathbf{r}_e,t)$ is neglected. In dipole approximation, the retention of electron's coordinate relative to the a.c.m position is stipulated by the prohibition of two-photon emission processes for the scheme under discussion.

To consider the effect of pump light on a two-level atom we will use the "dressed state" picture [6]. In accepted here classical field representation these dressed states are two quasi-stationary solutions of Schrödinger equation with Hamiltonian $\hat{H}_0 - \hat{\mathbf{d}}\mathbf{E}_p(z,t)$ in the rotating wave approximation. The characteristic feature of these states is the constant in time probability of populating atomic level $1$ (or $2$). Explicitly, the atomic wavefunction corresponding to each of these solutions is given as [6,7]

$$|\Psi_\pm(z,t)\rangle = N_\pm \left[|1\rangle - \frac{2\lambda_\pm}{\Omega}|2\rangle \exp(-i\omega_p t + ik_p z)\right] \exp\left(-\frac{i}{\hbar}(E_1 + \hbar\lambda_\pm)t\right). \qquad (5)$$

Here by $\Omega = 2\mathbf{d}\mathbf{E}_p/\hbar$ and $\Omega' = \sqrt{\Delta^2 + \Omega^2}$ are denoted the familiar Rabi and generalized Rabi frequencies respectively, $\Delta = \omega_p - \omega_0$ denotes the resonance detuning, $\lambda_\pm = -\Delta/2 \pm \Omega'/2$ stands for the high-frequency Stark shift of energy levels, and $N_\pm = \Omega/\left[2\Omega'(\Omega' \mp \Delta)\right]^{1/2}$ is the normalization factor.

Assuming that without the weak radiation field atoms are in a superposition of dressed states (with known probability amplitudes α and β respectively), the state vector of an interacting atom may be written as:

$$|\Psi(\mathbf{r},t)\rangle = (\alpha + C_+(\mathbf{r},t))|\Psi_+(z,t)\rangle + (\beta + C_-(\mathbf{r},t))|\Psi_-(z,t)\rangle, \qquad (6)$$

where the sought amplitudes $C_+(\mathbf{r},t)$ and $C_-(\mathbf{r},t)$ represent perturbations due to interaction with weak (signal and idler) radiated field (2). For the quantum mechanical average of dipole moment operator $\langle \hat{d} \rangle_w \equiv \langle \Psi(\vec{r},t) | \hat{d} | \Psi(\vec{r},t) \rangle_w$ we will get

$$\langle \hat{\mathbf{d}} \rangle_w = \left(\alpha^* C_+ + \alpha C_+^*\right) \langle \Psi_+ | \hat{\mathbf{d}} | \Psi_+ \rangle + \left(\alpha^* C_- + \beta C_+^*\right) \langle \Psi_+ | \hat{\mathbf{d}} | \Psi_- \rangle \\ + \left(\alpha C_-^* + \beta^* C_+\right) \langle \Psi_- | \hat{\mathbf{d}} | \Psi_+ \rangle + \left(\beta^* C_- + \beta C_-^*\right) \langle \Psi_- | \hat{\mathbf{d}} | \Psi_- \rangle. \tag{7}$$

In the above expression, we omit the second order terms, relative to perturbation amplitudes $C_+(\mathbf{r},t)$ and $C_-(\mathbf{r},t)$. We also should switch from dipole matrix elements in dressed state basis $\langle \Psi_\pm | \hat{\mathbf{d}} | \Psi_\pm \rangle$ to matrix elements in bare state basis $\langle 1 | \hat{\mathbf{d}} | 2 \rangle \equiv d_{12} = d_{21}^*$, viz-namely, to the main optical characteristic of the optical transition $|1\rangle - |2\rangle$. Then, in order to complete the calculation of $\langle \hat{\mathbf{d}} \rangle_w$ and express the right hand side of Eq.(1) via the sought weak field amplitudes $\mathbf{A}_s$ and $\mathbf{A}_i$, we have to calculate $C_\pm(\mathbf{r},t)$s as functions of these amplitudes, solving to this end the Schrödinger equation with the Hamiltonian (3) and wave function (6). Calculations are carried out in the frame of standard perturbation (relative to weak field interaction potential (4)) theory and yield in a rather long expression, which we see inexpedient to bring here. Simultaneously, we neglect the second order derivatives of slowly varying amplitudes $\mathbf{A}_s$ and $\mathbf{A}_i$ on the left-hand side and the first order temporal derivatives on the right-hand side of Eq.(1). Further, we restrict ourselves by the rotating wave approximation, neglecting in the right-hand side of the equation all the terms not alternating in phase with the left-hand side of equation. Since we consider nondegenerate regime of generation, this procedure splits the equation into a pair of equations. Each one of them on its left-hand side contains only one of the weak field amplitudes, $\mathbf{A}_s$ or $\mathbf{A}_i$. As to the right-hand side terms, they can be divided into three groups for each of the obtained equations. The process of interest, 3-photon PDC, is determined by one of them. The second one gives the contribution to the value of gas refractive index. This contribution is, however, too small for conditions under consideration, and we neglect it. The third group of terms represents the well-known 4-photon parametric generation [7,8], where two pump photons are transforming into a pair of signal and idler photons ($2\omega_p \to \omega_s + \omega_i$). The frequency domain of this process lies near the pump frequency and the generated here frequencies about two times surpass the frequencies of 3-photon PDC. Taking into account also the absence of any back action of parametrically generated fields on the pump (source) field in frame of the perturbation theory, we can regard these two parametric processes as mutually independent. Putting aside the terms responsible for the 4-photon parametric process, we arrive to an "intermediate" point, i.e., to a sought pair of one-dimensional reduced wave equations, determining the 3-photon generation process of interest.

It is worth noting that in the adopted model of two level atoms with definite spatial parities of corresponding eigenstates, the $\mathbf{A} \cdot \hat{\mathbf{p}}$- term in potential (4) does not have any contribution in the generation process of interest in dipole, as well as higher approximations. As to contribution of the $\mathbf{A}_p \mathbf{A}_w$-term, it has three components. One, satisfying the condition $\omega_s + \omega_i = \omega_p$, and two additional components, red- and blue-sideband, which are frequency shifted relative to the ordinary component by minus or plus $\Omega$ respectively. Assuming the signal registration time much longer than the time of Rabi oscillation $2\pi/\Omega$, we can consider the mentioned components as independently processing ones.

The obtained equation for the generated field has the form:

$$\frac{d\mathbf{A}_s}{dz} = \frac{2\pi n e^2}{m c \hbar \omega_p}\left(|\alpha|^2 - |\beta|^2\right)\frac{\Delta}{\sqrt{\Delta^2 + \Omega^2}} \langle 1|\sin(k_s \rho_z)|2\rangle \left(\frac{d_{21}}{2\omega_p - \omega_i} + \frac{d_{12}}{\omega_i}\right)\left(\mathbf{E}_p \cdot \mathbf{A}_i^*\right), \qquad (8)$$

The interchange of indexes $s \leftrightarrow i$ in (8) will give the second sought wave equation. Here $\rho_z$ denotes the z-projection of electron's radius vector $\boldsymbol{\rho}$ relative to a.c.m. ($\mathbf{r}_e = \mathbf{r} + \boldsymbol{\rho}$). We have also used a well satisfied condition $\omega_{s,i} \gg \Omega$, to simplify, in some extend, the right-hand side expression in (8).

Now, neglecting the wave polarization effects (all the waves have the same linear polarization), we arrive to an elementary solution

$$A_{s,i}(z) = A_{s,i}(0)\exp(\alpha_0 z), \qquad (9)$$

where the coefficient of amplification is

$$\alpha_0 = \frac{4\pi n e^2}{m c}\left||\alpha|^2 - |\beta|^2\right|\frac{|\Delta|\Omega}{\sqrt{\Delta^2 + \Omega^2}} \sqrt{\frac{\langle 1|\prod_{\mu=s,i}\sin(k_\mu \rho_z)|1\rangle}{\prod_{\mu=s,i}(2\omega_p - \omega_\mu)\omega_\mu}} . \qquad (10)$$

Just like in ordinary balance equations, the value of amplification coefficient depends on the difference of state populations. Mutual coherency of dressed states has no role for this component of PDC.

Under the condition $\omega_s + \omega_i = \omega_p + \Omega$ we will get the blue-sideband component of generation. It is governed by the following reduced wave equation

$$\frac{dA_s}{dz} = \frac{2\pi n e^2}{mc\hbar\omega_p} \alpha^*\beta \frac{\Omega}{\Delta^2 + \Omega^2} \langle 1|\sin(k_s\rho_z)|2\rangle \left(\frac{\Omega' d_{21}}{2\omega_p - \omega_i} + \frac{\Delta d_{12}}{\omega_i}\right)\left(E_p \cdot A_i^*\right) e^{i\Omega z/c} \quad (11)$$

and its conjugate equation, with $s \leftrightarrow i$ index interchange. For the corresponding coefficient of amplification (without polarization peculiarities), we obtain

$$\alpha_\Omega = \frac{\pi n e^2}{mc\omega_p} |\alpha^*\beta| \frac{\Omega^2}{\Delta^2 + \Omega^2} \sqrt{\langle 1|\prod_{\mu=s,i}\sin(k_\mu\rho_z)|1\rangle \prod_{\mu=s,i}\left(\frac{\Omega'}{2\omega_p - \omega_\mu} + \frac{\Delta}{\omega_\mu}\right)}. \quad (12)$$

Being proportional to the product term $|\alpha^*\beta|$, this generation is conditioned by the superposition nature of atomic states and, in contrary to the ordinary component, is sensitive towards the mutual decoherency of the superimposed dressed atomic states. This follows from the fact, that in classical picture field amplitudes are observable quantities. Therefore, in case under consideration, if there is some noise in the initially prepared state, the averaging must be carried out in the wave equation Eq.(11). Note, that the envisioned amplification gets its maximum value at point $\alpha = \beta$, where the ordinary, balance-type component of PDC is completely canceled from the amplification process.

What is even more remarkable in this case is that the amplification coefficient (12) continuously increases with the pump field intensity $\Omega$, being almost linear in the asymptotic range $\Omega \gg |\Delta|$. In other words, this coefficient does not exhibit a property of saturation, which was typical in ordinary case of generation (see Eq.(10)). It is also important to note that for a common intensity $\Omega \approx |\Delta|$, the coefficient $\alpha_\Omega$ is already of the order of $\alpha_0$.

The red-sideband component of the PDC generation processes at the condition $\omega_s + \omega_i = \omega_p - \Omega$ and is described by the reduced wave equation

$$\frac{dA_s}{dz} = \frac{2\pi n e^2}{mc\hbar\omega_p} \alpha^*\beta \frac{\Omega}{\Delta^2 + \Omega^2} \langle 1|\sin(k_s\rho_z)|2\rangle \frac{(2\Delta + \Omega') d_{21}}{2\omega_p - \omega_i} \left(E_p \cdot A_i^*\right) e^{-i\Omega z/c} \quad (13)$$

and its conjugate one with $s \leftrightarrow i$ index interchange. The corresponding amplification coefficient is

$$\alpha_{-\Omega} = \frac{\pi n e^2}{m c \omega_p} |\alpha^* \beta| \frac{\Omega^2}{\Delta^2 + \Omega^2} \frac{|\Delta + \Omega/2|}{\sqrt{(2\omega_p - \omega_s)(2\omega_p - \omega_i)}} \sqrt{\langle 1| \prod_{\mu=s,i} \sin(k_\mu \rho_z) |1 \rangle}. \tag{14}$$

As we see, $\alpha_{-\Omega}$ too, just as the "mirror" coefficient $\alpha_\Omega$, has a nonsaturating character for the dependence on pump wave intensity $\Omega$, but the frequency dependencies for the sideband amplification coefficients (14) and (12) are somewhat different.

### 3. Summary

In this paper we suggest a new possibility for a gas-phase PDC process, which entirely originates from the superposition nature of the internal atomic state. It appears as generation of blue- and red-sideband components, relative to the ordinary one, and occurs for frequencies satisfying the conditions $\omega_s + \omega_i = \omega_p \pm \Omega$ respectively. The "raisin" of this new process is its nonsaturability, when the pump wave intensity is increased. As to the amplification of the ordinary component, it behaves as usually and saturates in the limit of high intensities.

It should be noted that in the chosen model frame the presented theory is exact with respect to pump field intensity. We assumed fulfillment of rotating wave approximation, resonance approximation, and, of course, a two-level structure for the atomic energy spectrum. We don't implicate a request of high degree of mutual coherency between the populated dressed states, since it enters into the problem of interaction as an initial condition. As is seen, the range of validity for generation of nonsaturable PDC sidebands is quite broad. The feeblest point here seems to be the two-level model. Nonsaturable behavior in PDC will be destroyed by perceptible population of other energy levels, and/or if the coupled energy levels will have a.c. Stark shifts comparable to the distance to the closest energy level.

The given below analysis renders the simplest two-level scheme of PDC process subject to experimental observation. For illustrative purposes of this opportunity, we assume ordinary dipole-allowed optical transition with induced dipole moment of $|\mathbf{d}_{12}| = 10^{-17}$ CGSE and wavelength $\lambda_0 = 2\pi c/\omega_0 \simeq 1\,\mu m$. Taking typical atomic size $\bar{\rho} = 3\,\text{Å}$, resonance detuning $\Delta = 10\,\text{GHz}$ and the central part of down converted frequencies, and assuming the atomic number density to be on the order of atmospheric density, one arrives to $\alpha_\Omega = \alpha_{-\Omega} = 10^{-3}\,\text{cm}^{-1}$ for the light flux of $1\,\text{kW}/\text{cm}^2$, with a perspective of almost linear enhancement for larger field strengths. We also note that the amplification rate may significantly vary depending, e.g., on the electron's orbit effective radius of relevant energy levels.

This work was supported by the A. von Humboldt Foundation, by the NFSAT/CRDF Grant № UCEP-0207 and by the Armenian Science Ministry Grant № 143.